\documentclass[twocolumn,aps,prl,showpacs]{revtex4}

\usepackage{graphicx}

\begin{document}

\title{Robust half-metallic ferromagnet of cubic VGe$_3$Si$_4$}

\author{San-Dong Guo and Bang-Gui Liu}
\affiliation{Institute of Physics, Chinese Academy of Sciences,
Beijing 100190, China} \affiliation{Beijing National Laboratory
for Condensed Matter Physics, Beijing 100190, China}

\begin{abstract}
To seek half-metallic ferromagnets compatible with the GeSi
semiconductor, we use state-of-the-art density-functional-theory
methods to study cubic $\mathrm{VGe_3Si_4}$. After optimizing its
crystal structure with various magnetic orders considered, we find
the ferromagnetic structure is the ground state phase and
investigate its stability, and then study its spin-resolved
electronic structure with an improved method. Our calculated
results show that the cubic $\mathrm{VGe_3Si_4}$ is a
half-metallic ferromagnet with half-metallic gap 350 meV, and it
is structurally stable against deformations and magnetically
robust against antiferromagnetic fluctuations. These results make
us believe that the cubic $\mathrm{VGe_3Si_4}$ could be
synthesized as good thin films soon and used in practical
spintronic devices.
\end{abstract}

\pacs{75.30.-m,71.20.-b,75.50.-y,75.90.+w}

\maketitle

Half-metallic ferromagnetic materials can play important roles in
high-performance spintronic applications because their special
electronic structures can supply high spin-polarization at high
temperatures \cite{spintr1,spintr2}. This kind of materials are
special in the sense that one of their two spin channels is
metallic and the other nonmetallic, which causes a
spin-polarization of nearly 100\% (there is a tiny, or small,
modification due to the spin-orbit coupling effect) at the Fermi
energy \cite{hm}. NiMnSb, CrO$_2$, Fe$_3$O$_4$, and Co$_2$MnSi are
typical examples for such materials
\cite{hm,cro2,fe3o4,co2mnsi,hmsp}. Because of the central roles of
semiconductor technology in modern electronic devices, it has been
important to seek high-performance half-metallic ferromagnetic
materials compatible with important semiconductors
\cite{nhm,mnas,mnas1,cras,crsb,crassurf,lbg1,lbg2,zbtm56,pickett6,lbg3,sanyal,crte,gsd}.
Great advance has been achieved in transition-metal substituted
III-V and II-VI semiconductors, such as CrAs, CrSb, MnSb, and CrTe
\cite{cras,crsb,crassurf,lbg1,lbg2,zbtm56,pickett6,lbg3,sanyal,crte}.
For semiconductor spintronic applications, it is of much interest
to explore half-metallic ferromagnets based on the important
group-IV semiconductors such as Si, Ge, and GeSi.

In this Letter we use state-of-the-art full-potential
density-functional-theory methods to study structural, electronic,
and magnetic properties of cubic $\mathrm{VGe_3Si_4}$ based on the
GeSi semiconductor. We optimize its crystal structure with various
magnetic orders considered, and then investigate the structural
and magnetic stability of the ferromagnetic structure as the
ground state phase and study its spin-resolved electronic
structure with an improved method. Our calculated results show
that the cubic $\mathrm{VGe_3Si_4}$ is a half-metallic ferromagnet
with half-metallic gap 0.35 eV, and it is structurally stable
against deformations and magnetically robust against
antiferromagnetic fluctuations. More detailed results will be
presented in the following.

We use a full-potential linearized augmented-plane-waves method
within the density functional theory \cite{dft}, as implemented in
package WIEN2k \cite{wien2k}. We use the generalized gradient
approximation\cite{pbe96} to optimize the structures and calculate
all the total energies, but we use a modified Becke-Johnson exchange
potential \cite{mbj} to calculate the final density of states and
energy bands to be presented because this method has been proved to
give much better semiconductor gaps for various semiconductors and
insulators. Full relativistic effects are calculated with the Dirac
equations for core states, and the scalar relativistic approximation
is used for valence states \cite{relsa,relsa1,relsa2}. The
spin-orbit coupling is neglected because it has little effect on our
results. We use 2000 k points in the first Brillouin zone, make
harmonic expansion up to $l_{\rm max}=10$ in the atomic spheres, and
set $R_{mt}*k_{\rm max}=7$. The radii of the atomic spheres of V,
Ge, and Si are set to 2.27, 2.16, and 2.01 atomic unit,
respectively. The volumes are optimized in terms of total energy
method, and the internal atomic position parameters with a force
standard of 3 mRy/a.u. The self-consistent calculations are
considered to be converged when the integration of the absolute
charge-density difference between the input and out density is less
than 0.0001$|e|$ per formula unit, where $e$ is the electron charge.

We construct the cubic $\mathrm{VGe_3Si_4}$ by substituting V for
the corner Ge atom in the unit cell of the crystalline GeSi with
zincblende structure, as is shown in Fig. 1. The resultant structure
has space group No. 215. We optimize its lattice constant and
internal atomic position parameters. The optimized lattice constant
equals 5.632\AA{} and the internal Si position parameter is 0.2509.
Its total magnetic moment is 1.000$\mu_B$ per formula unit, can be
attributed to the 3$d$ electron of the V$^{4+}$ ion. The
$\mathrm{VGe_3Si_4}$ is a half-metallic ferromagnet, but its
minority-spin conduction band bottom almost touches the Fermi level.
In order to improve the results, we re-do the self-consistent
calculation, with all the structural parameters, using an improved
exchange-correlation method including the modified Becke-Johnson
potential because this new method improves semiconductor gaps and
$d$-state positions for various materials \cite{mbj}. We present the
density of states between -5 eV and 3 eV in Fig. 2. It is clear that
the majority-spin channel is metallic and the minority-spin channel
has an energy of 0.93 eV across the Fermi level, and therefore the
cubic $\mathrm{VGe_3Si_4}$ is a half-metallic ferromagnet with
half-metallic gap 350 meV \cite{lbgb}. The spin-resolved bands
between -8 eV and 4 eV are presented in Fig. 3. The two
majority-spin bands across the Fermi level are from V 3$d$-$e_g$
doublet, and the lower bands in the energy window can be attributed
to the $p$ and $s$ states from Ge and Si. The spin exchange
splitting should be driven by the V $d$ electrons. The minority-spin
gap of 0.93 eV across the Fermi level, and hence the
half-metallicity, is due to both the spin exchange splitting and the
right position of V $d$ states in the GeSi gap.

\begin{figure}[tb]
\includegraphics[width=5cm]{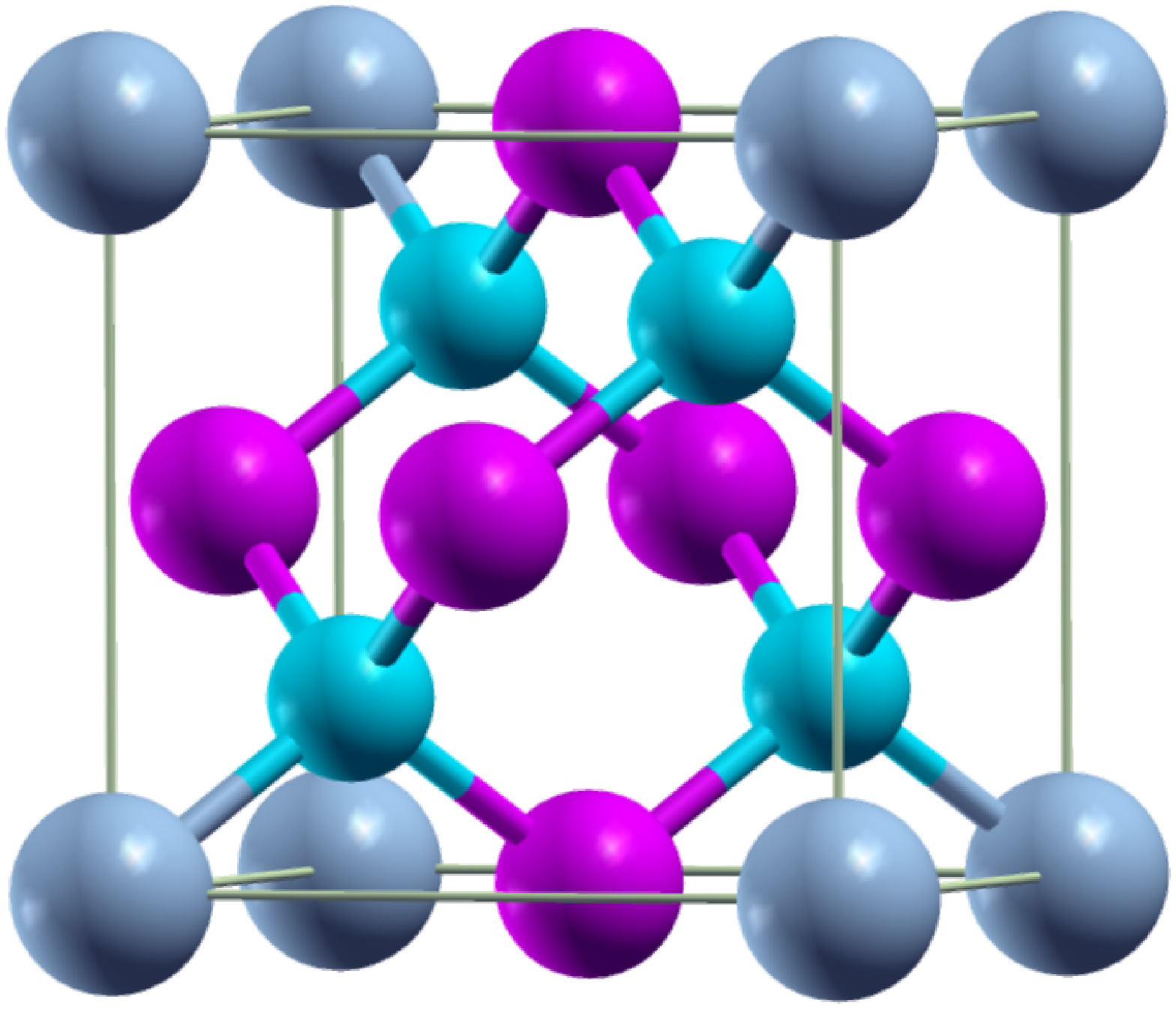}
\caption{(color online) The crystalline unit cell of the cubic
$\mathrm{VGe_3Si_4}$ with space group No. 215. The largest balls
(corner) represents V atoms, the medium (face-center) Ge, and the
smallest (in the cube) Si. }
\end{figure}

\begin{figure*}[!htb]
\includegraphics[width=12cm]{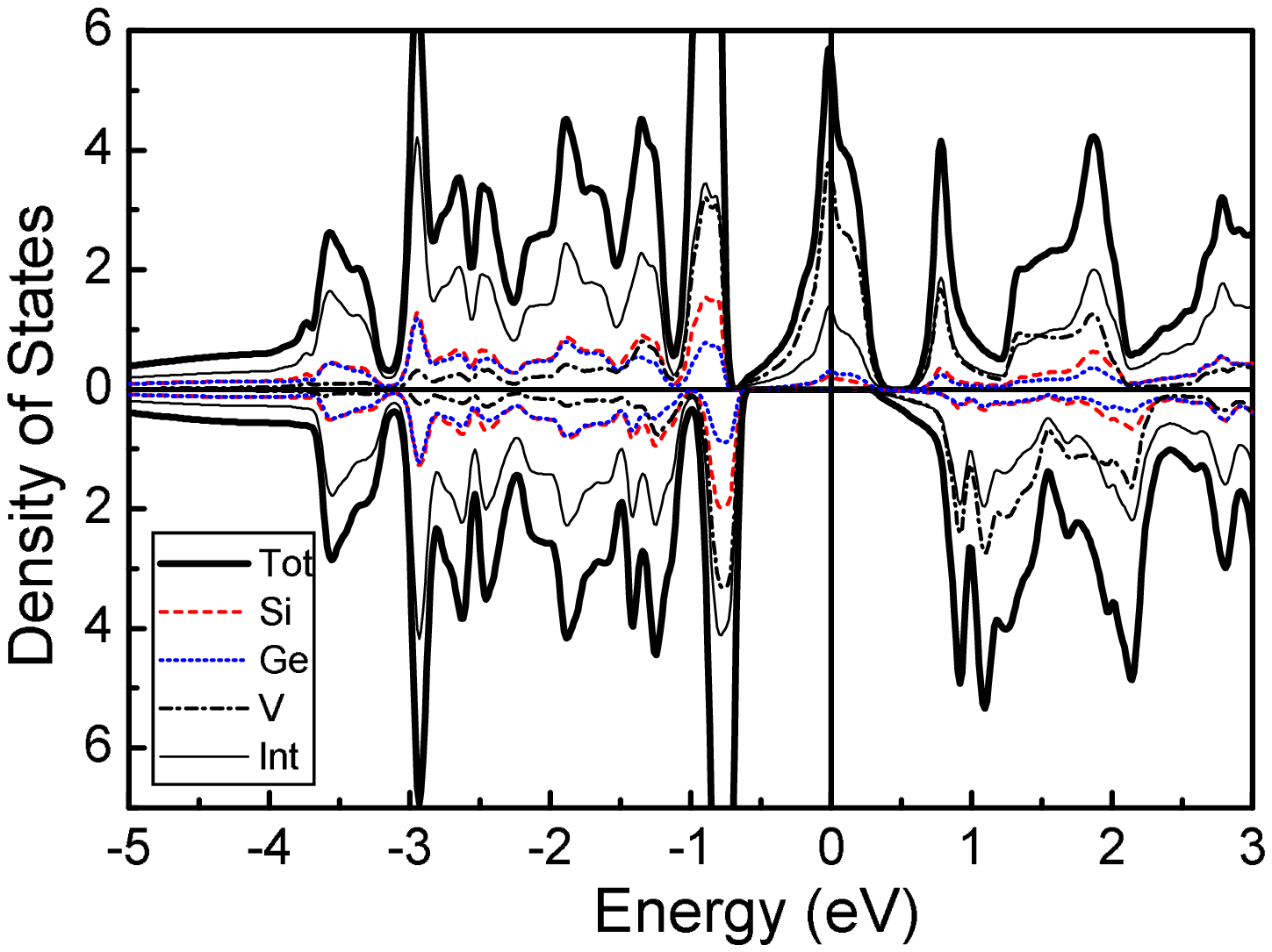}
\caption{(color online) Spin-dependent density of states (DOS, in
state/eV per formula unit) of the ferromagnetic
$\mathrm{VGe_3Si_4}$. The solid thick line represents the total DOS,
and the dashed, dotted, dot-dashed, and thin solid lines describe
the partial DOSs projected in the Si, Ge, and V atom spheres and the
interstitial region, respectively.}\label{dos}
\end{figure*}

\begin{figure*}[!htb]
\includegraphics[width=7cm]{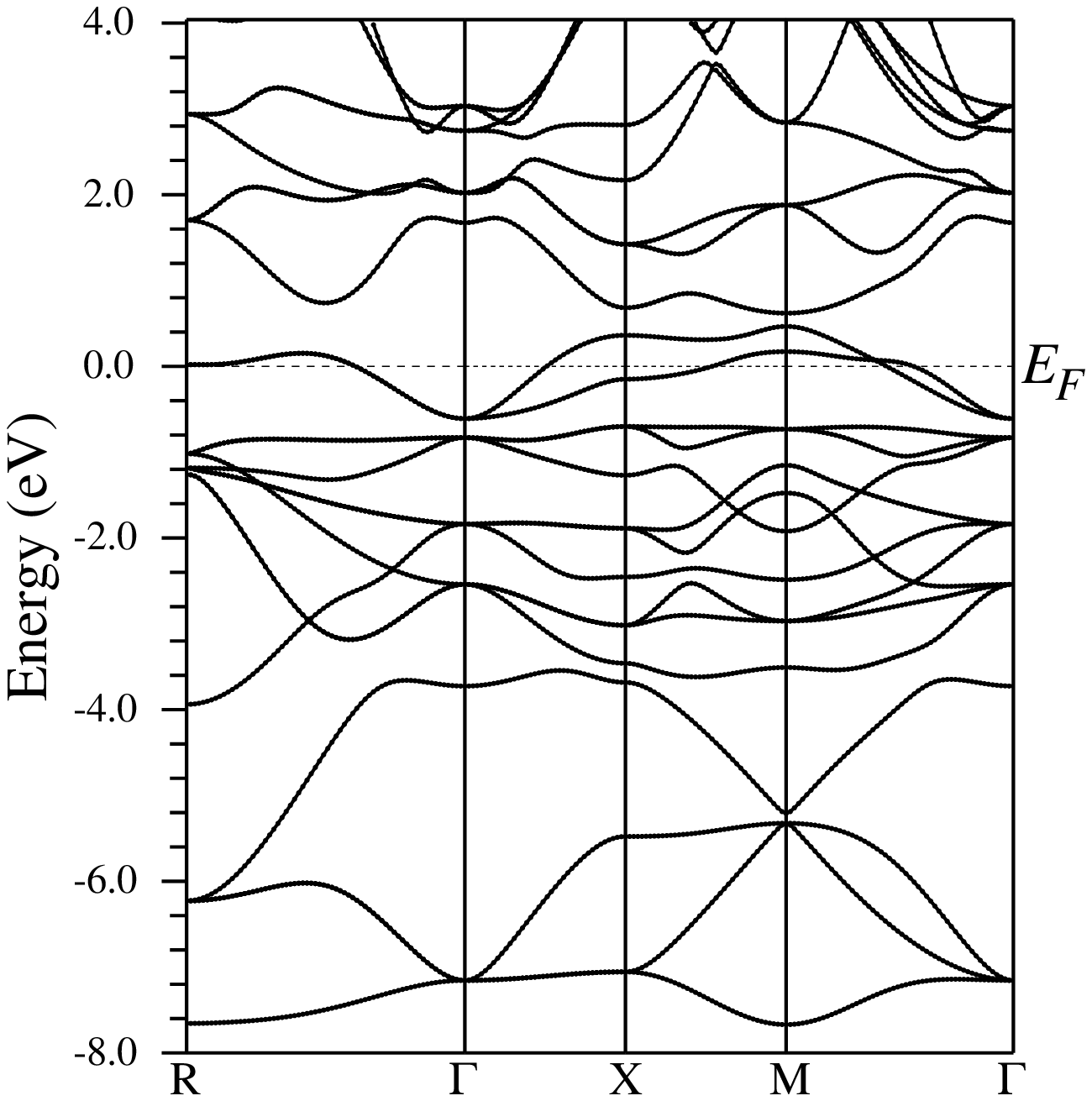}
\includegraphics[width=7cm]{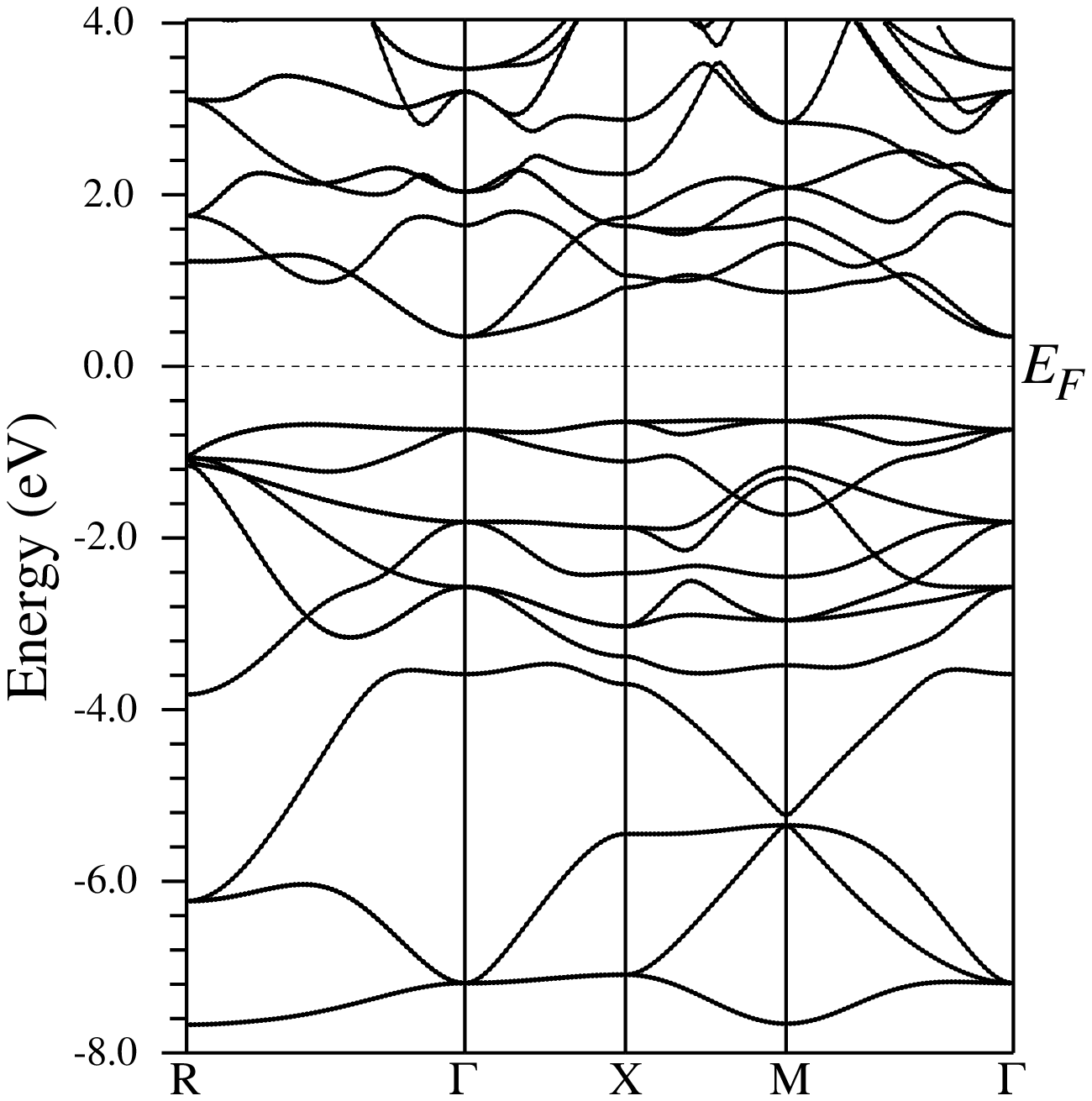}
\caption{Spin-dependent energy bands of the ferromagnetic
$\mathrm{VGe_3Si_4}$. The left-side panel shows the majority-spin
bands and the right-side one the minority-spin bands.}\label{eb}
\end{figure*}

To investigate the robustness of the ferromagnetic order against
possible antiferromagnetic fluctuations, we calculate the total
energies for nonmagnetic structure and three kinds of
antiferromagnetic configurations, namely AFG (0.5,0.5,0.5)$\frac{2
\pi} {a}$, AFA (0,0,0.5)$\frac{2 \pi} {a}$, AFC
(0.5,0.5,0)$\frac{2 \pi} {a}$. They are constructed by doubling
the unit cell along the [111], [110], and [001] directions,
respectively. Our results show that the nonmagnetic structure is
155 meV higher in total energy per formula unit than the
ferromagnetic structure, and the three antiferromagnetic
structures are 49, 51, and 60 meV higher in total energy per
formula unit, respectively. Therefore, the ferromagnetic order of
the cubic $\mathrm{VGe_3Si_4}$ is indeed stable against
antiferromagnetic fluctuations, and these energy differences
indicate that a room-temperature should be reached for the Curie
temperature. Since the half-metallic ferromagnets from II-VI and
III-V semiconductors have small values for the elastic modulus
$C^\prime$, usually smaller than 10 GPa, it is important to
calculate the elastic moduli, $B$, $C^\prime$, and $C_{44}$, for
the cubic $\mathrm{VGe_3Si_4}$. All these calculated results are
summarized in Table \ref{tt}. It is impressive that the $C^\prime$
value, the smallest among the three moduli, equals 23.3 GPa,
substantially larger than those of the half-metallic ferromagnets
based on II-VI and III-V semiconductors. This means that much
better thin-film quality can achieved and therefore there will be
a much better potential for high-performance device applications.

\begin{table}[!htb]
\caption{The lattice constant ($a$), the Si position parameter
($z_x$), the total moment ($M_{\rm tot}$), the half-metallic gap
($G_{\rm HM}$), the elastic constants ($B$, $C^\prime$, $C_{44}$)
of the stable ferromagnetic structure, and the total energies of
the three antiferromagnetic (AFG, AFC, AFA) and the nonmagnetic
(NF) structures with respect to the ferromagnetic structure (FM).}
\begin{tabular*}{0.48\textwidth}{@{\extracolsep{\fill}}ccccc}
\hline\hline $a$ & $z_x$ & $M_{\rm tot}$ & $G_{\rm HM}$ & \\
5.632 \AA{} & 0.2509 & 1.000 $\mu_B$ & 0.35 eV & \\ \hline
$B$ & $C^\prime$ & $C_{44}$ & & \\
70.9 GPa & 23.3 GPa & 65.8 GPa & & \\ \hline FM & AFG & AFC & AFA & NF\\
0 & 49. meV & 51. meV & 60. meV & 155. meV\\ \hline\hline
\end{tabular*}\label{tt}
\end{table}

In summary, we find that the cubic $\mathrm{VGe_3Si_4}$ is a
half-metallic ferromagnet with a half-metallic gap 350 meV. We
have optimized its crystal structure and investigated its crystal
structure against deformations and the robustness of its
ferromagnetic order against antiferromagnetic fluctuations. Our
results show that its ferromagnetic order is stable, maybe enough
to keep a Curie temperature at room temperature, and $C^\prime$,
the smallest among the three elastic moduli,  reaches 23.3 GPa,
substantially larger than those of the half-metallic ferromagnetic
materials based the II-VI and III-V semiconductors. These results
make us believe that the cubic $\mathrm{VGe_3Si_4}$ could be
synthesized as thin films of enough thickness soon in the near
future and used in spintronic applications.

\begin{acknowledgments}
This work is supported by Nature Science Foundation of China
(Grant Nos. 10874232 and 10774180), by the Chinese Academy of
Sciences (Grant No. KJCX2.YW.W09-5), and by Chinese Department of
Science and Technology (Grant No. 2005CB623602).
\end{acknowledgments}

\end{document}